\documentclass[twocolumn,showpacs,amssymb,aps,nofootinbib,floatfix]{revtex4}

\usepackage{amsmath}
\usepackage{epsfig,color}

\begin{document}
\title{The Kolmogorov-Smirnov test and its use for the identification of fireball fragmentation}
\author{Ivan Melo$^a$, Boris Tom\'a\v{s}ik$^{b,c}$, Giorgio Torrieri$^{d}$, Sascha Vogel$^{e}$,\\ 
Marcus Bleicher$^{e}$, Samuel Kor\'ony$^{b}$, Mikul\'a\v{s} Gintner$^{a,b}$}
\affiliation{$^a$ \v{Z}ilinsk\'a Univerzita, Univerzitn\'a 1, 01026 \v{Z}ilina, Slovakia\\ 
$^b$ Univerzita Mateja Bela, Tajovsk\'eho 40, 97401 Bansk\'a Bystrica, Slovakia\\
$^c$ Faculty of Nuclear Science and Physics Engineering, Czech Technical University in Prague, 
B\v{r}ehov\'a 11, 11519 Prague, Czech Republic\\
$^d$ Frankfurt Institute of Advanced Studies, Johann Wolfgang Goethe Universit\"at, 
Ruth-Moufang-Str.\ 1, 60438 Frankfurt am Main, Germany\\
$^e$ Institut f\"ur theoretische Physik, Johann Wolfgang Goethe Universit\"at,
Max-von-Laue-Str. 1, 60438 Frankfurt am Main, Germany}

\date{February 10, 2009}

\begin{abstract}
We propose an application of the Kolmogorov-Smirnov test for  rapidity 
distributions of individual events in ultrarelativistic heavy ion collisions. 
The test is particularly suitable to recognise non-statistical differences 
between the events. Thus when applied to a narrow centrality class it could 
indicate differences between events which would not be expected if all events
evolve according to the same scenario. In particular, as an example we assume here a possible 
fragmentation of the fireball into smaller pieces at the quark/hadron phase 
transition. Quantitative studies are performed with a Monte Carlo model capable 
of simulating such a distribution of hadrons. We conclude that the Kolmogorov-Smirnov 
test is a very powerful tool for the identification of the fragmentation process.
\end{abstract}
\maketitle


\section{Introduction}

The highly excited matter created in ultrarelativistic nuclear collisions expands very fast. It is commonly 
accepted that a deconfined phase has been reached in Au+Au collisions at RHIC 
\cite{whitepaperBRAHMS,whitepaperPHOBOS,whitepaperSTAR,whitepaperPHENIX}, while 
the  onset of deconfinement has been advocated at SPS energies \cite{na49onset,cernpress}.

While in lattice QCD calculations a static thermodynamic medium is assumed, in heavy ion collisions
the situation is vastly different. Here, the longitudinal expansion dynamics leads to a rapid 
passage from the deconfined to the confined phase.
A system which undergoes the phase transition quickly may not follow the usual equilibrium
scenario. In fact, for a first-order phase transition, the high temperature phase may survive down to 
temperatures drastically below the transition temperature, i.e.\ the system supercools. If the expansion rate 
is faster than the nucleation rate of bubbles of the new phase, the system reaches the point 
of spinodal instability\footnote%
{This is the inflection point of the dependence of entropy on an extensive variable, 
see e.g. \cite{Chomaz:2003dz,Randrup:2005kp}.
}.
Beyond such a point, entropy is gained if the system separates into two phases and so it becomes 
mechanically unstable. Spinodal fragmentation connected with nuclear liquid/gas phase transition 
has been identified in heavy ion collisions at few hundred MeV per nucleon \cite{Chomaz:2003dz,Randrup:2005kp}, and it has 
been proposed that it might be the actual scenario at ultrarelativistic energies as well 
\cite{Mishustin:1998eq,Scavenius:2000bb}. Fragmentation assumes that at the phase transition 
the system decays into droplets of smaller size. These droplets then emit hadrons.

Lattice calculations indicate, however, that at RHIC and LHC the transition from partonic to 
hadronic matter is a rapid but smooth crossover \cite{Aoki:2006we}. 
Thus spinodal decomposition seems irrelevant scenario in this 
case. On the other hand,  conformal symmetry is broken
close to the phase transition and as a consequence the \emph{bulk} viscosity---being negligible 
otherwise---shows a peak here \cite{kharbulk,pratt,Torrieri:2007fb}. Bulk viscosity acts against the expansion and 
slows it down. As a result, if the system previously accumulated kinetic energy due to 
expansion, it may fragment \cite{Torrieri:2007fb}. An analysis of hydrodynamic instabilities 
shows that such a scenario may be realistic  \cite{Torrieri:2008ip}.

Hence, it appears that fragmentation may happen in ultrarelativistic nuclear collisions. 
Many kinds of observables might be sensitive to it. Most notable are multiplicity 
fluctuations in varying rapidity windows \cite{Baym:1999up}, fluctuations of mean $p_t$ 
\cite{Broniowski:2005ae}, rapidity correlations \cite{Pratt:1994ye,Randrup:2005sx}, proton and
kaon correlations \cite{Pratt:1994ye}, $\phi$-meson production \cite{Pratt:1994ye}, and
two-pion femtoscopy \cite{granularHBT,Torrieri:2007fb,granularimaging}. 

In this paper we inspect  event-by-event fluctuations of rapidity distributions. 
If final state hadrons are emitted from droplets, their velocities will be close to those of the droplets.
Thus, clustering would appear in their momentum distribution. Moreover, in each event clusters will have
different velocities. An important 
contribution to clustering will also come from the resonance decays and we shall investigate this effect. 
Thus, the momentum distribution will vary from event to event. Specifically, here we shall compare
\emph{rapidity distributions} from different events and look for differences due to fireball fragmentation. 
To this end, we employ the Kolmogorov-Smirnov (KS) test which can be used for identification 
of non-statistical differences between two empirical distributions \cite{kolm,smir}. An important advantage 
of the KS test is its independence from the underlying distribution of the measured quantity. 

In the next Section we shortly introduce the KS test. Then, we illustrate its sensitivity by 
a couple of toy simulations. For more realistic studies we generate artificial data with the help 
of the event generator DRAGON \cite{Tomasik:2008fq} which is very briefly introduced in Section 
\ref{quag}. Results obtained with these data are presented in Section \ref{results}. We conclude 
in Section \ref{conc}. In the Appendix we review the evaluation of the cummulative distribution function 
for the Kolmogorov distribution.


\section{The Kolmogorov-Smirnov test and how to use it}
\label{KStest}

Let us start by explaining the technical part of the problem. One has two empirical distributions in 
variable $x$, which can be rapidity, $p_t$, or yet something else\footnote{
Note that for a cyclic variable, e.g.\ azimuthal angle $\phi$, the KS test cannot be applied and 
instead a modification known as Kuiper test \cite{kuiper,stephens65}must be employed.
}.
(In the present work we work with rapidities.) 
The multiplicities may differ. The question we want to ask 
is, whether the two empirical distributions are the same in the sense that they would correspond to the 
same underlying theoretical single-particle probability density, and there are no correlations between 
particles in one event. 

\begin{figure}[t]
\begin{center}
\includegraphics[width=1.0\linewidth]{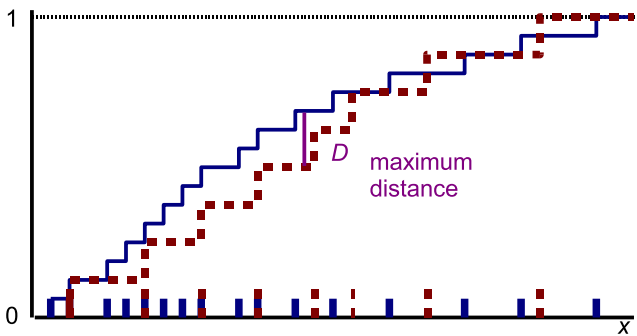}
\end{center}
\caption{
Construction of the two empirical cummulative distribution functions, one with thin solid (blue) line 
and one with thick dashed (red) line. The maximum distance between them is 
$D$.}
\label{f:KSD}
\end{figure}
Practically, the quantity $x$ is measured for each particle in an event. The empirical cummulative 
distribution function (ECDF) is constructed so that a step of the height 
$1/n_i$ ($n_i$ is the multiplicity of the event) is made on all positions of measured
$x$'s (Fig.~\ref{f:KSD}). This is done for both events of a pair. Subsequently, 
one finds the \emph{maximum} vertical distance between the two ECDF's and introduces 
\begin{equation}
d = \sqrt{n} D = \sqrt{\frac{n_1n_2}{n_1 + n_2}} D
\end{equation}
where $D$ is the distance of two ECDF's 
and $n_1$, $n_2$ are the multiplicities of the two data sets.
The procedure of the test is illustrated in Figure~\ref{f:KSD}.
The cummulative distribution function of the \emph{Kolmogorov distribution} concerns the case 
of events generated from the same underlying 
theoretical probability distribution for the quantity $x$, $\rho(x)$. It will be expressed with the help 
of the function $Q(d)$ as
\begin{equation}
P(d' < d) = 1-Q(d)\, .
\end{equation}
where $P(d'<d)$ is the probability that we find a difference $d'$ smaller than $d$. 
An important feature of this approach is that $Q(d)$ does not depend on the particular shape of the 
theoretical distribution $\rho(x)$.
Unfortunately, the general form of $Q(d)$ valid for any multiplicities and distances $d$ is not suitable 
for practical evaluation. 
Usable approximate expressions for $Q(d)$ are summarised in the Appendix. 
It follows that if many pairs of events would be drawn from a set of events generated all from  
the same underlying probablility  distribution and for each pair the quantities $d$ and
$Q(d)$ would be determined, then the $Q's$ would be distributed \emph{uniformly}. Deviation from 
uniform distribution, particularly an enhanced population of low $Q$'s (large $d$'s) 
indicates that the events are \emph{not drawn independently from the same underlying distribution}. 
In this way the KS test will be used here. 

Note that the KS test does not identify the physical origin of the difference between the events. It is
a robust way to identify that there is a difference, however, 
the origin must be singled out by other means. 
In addition to fireball fragmentation, 
these can be fluctuations of the initial state of the fireball 
evolution, final resonance decays, conservation laws, 
and quantum correlations. A detailed investigation of these will be pursued  in 
subsequent papers. The important message of the KS test is, that it can disprove the usual paradigm 
that data from many collisions (within the same centrality class) are produced by basically identical
fireballs.

\begin{figure}[t]
\begin{center}
\includegraphics[width=1.0\linewidth]{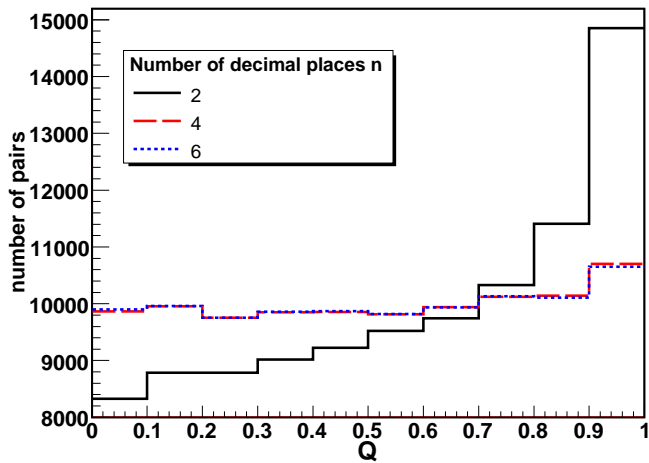}
\end{center}
\caption{
Histograms of $Q$'s from the KS test applied on $10^5$ pairs out of $10^5$ events generated 
from a uniform distribution between 0 and 1 in the variable $x$ (rapidity) and with multiplicities 
distributed according to Poisson distribution with the mean 200. Different histograms correspond 
to rapidity data truncated after 2, 4, and 6 decimal places. The histogram with 8 significant figures
is identical to that with 6 figures.}
\label{f:figures}
\end{figure}

It is important to realise that the number of significant 
decimal figures to which the quantity $x$ (rapidity here) 
is measured may also influence the result of the KS test if it is applied on a large number of pairs 
of events. This is illustrated in Figure~\ref{f:figures}.
The peak at $Q\to 1$ increases with lowering the number of decimal places taken into account. 
The explanation is trivial but instructive. Only rapidities between 0 and 1 were generated. Within 
$10^5$ events, each having multiplicity around 200, there are about $2\times 10^7$ particles. 
Hence, if their values of $x$ are given to less than 8 figures, we are guaranteed to have repeating 
values of $x$ in our sample. For 6 given figures we expect each value to appear on average 
20 times, for 4 figures it is 2000 times and for 2 figures we are even at 200,000 times! Clearly, 
this effect correlates the events since it artificially chooses $x$ from a finite number of possible 
values. According to its construction, the $D$'s will acquire smaller values on average. 

We have also checked that the normal fluctuation of the number of entries in a bin of the 
$Q$-histogram is equal to the square root of the number of entries. 
We thus propose the use of quantity
\begin{equation}
\label{Rdef}
R = \frac{N_0 - \frac{N_{\rm tot}}{B}}{\sigma_0} 
= \frac{N_0 - \frac{N_{\rm tot}}{B}}{\sqrt{\frac{N_{\rm tot}}{B}}}\, ,
\end{equation}
where $N_0$ is the number of pairs in the first bin of the $Q$-histogram (next to $Q=0$), 
$N_{\rm tot}$ is the total number of pairs,
$B$ is the number of bins of the $Q$-histogram, and $\sigma_0 = \sqrt{N_{\rm tot}/B}$ is the expected 
variance of the number of entries in the first bin. The modulus of $R$ should be of the order 
1; values considerably bigger than that indicate non-statistical differences between the events. 


\section{The sensitivity of the test}
\label{sens}

\begin{figure}[t]
\begin{center}
\includegraphics[width=1.0\linewidth]{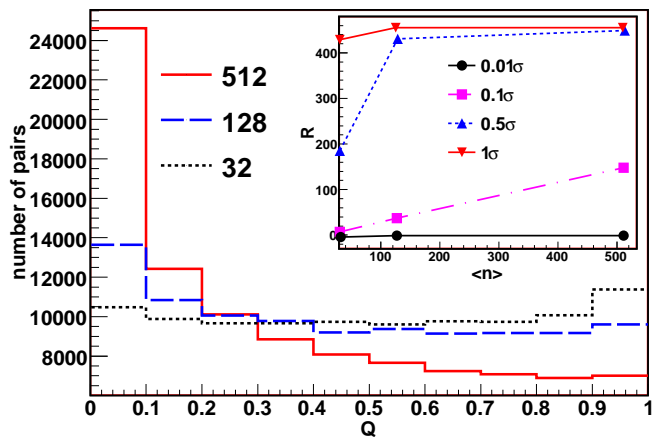}
\end{center}
\caption{(Color online)
Large plot: the $Q$ histograms from the KS test on event samples consisting from 
two classes of events. One class was generated from Gaussian distribution with the mean 0 and the width 
$\sigma=0.1$. The other class is generated from Gaussian distribution with the same width, 
but the mean is shifted by $0.1\sigma$. The multiplicities of the events are 32 (black dotted histogram), 
128 (blue dashed), and 512 (red solid). 
Smaller inset plot: the dependence of the parameter $R$ on the multiplicity of the events for
the difference of the means equal to $0.01\sigma$ (black circles), $0.1\sigma$ (magenta squares), 
$0.5\sigma$ (blue triangles), and $1\sigma$ (red upside down triangles).
}
\label{f:varmean}
\end{figure}

In this section we study how the proposed method works in case of clear cut examples. First, we generate 
samples  of ``events'' where one half of all events is generated according to Gaussian distribution
with the width $\sigma=0.1$. For the second half of events we keep the same width and vary the mean: we have samples 
with the mean shifted with respect to the other half by $2\sigma$, $1\sigma$, $0.5\sigma$, $0.1\sigma$, and $0.01\sigma$. 
In Figure~\ref{f:varmean}
we observe how the KS procedure recognises the difference of $0.1\sigma$ pretty well if the average multiplicity 
is 512, and how the resolution power decreases when lowering the multiplicity to 32. As seen in the 
smaller inset plot, for smaller distances between the two Gaussian means 
the difference is not recognised, while for larger distance the difference is resolved by the test 
for all multiplicities. 

\begin{figure}[t]
\begin{center}
\includegraphics[width=1.0\linewidth]{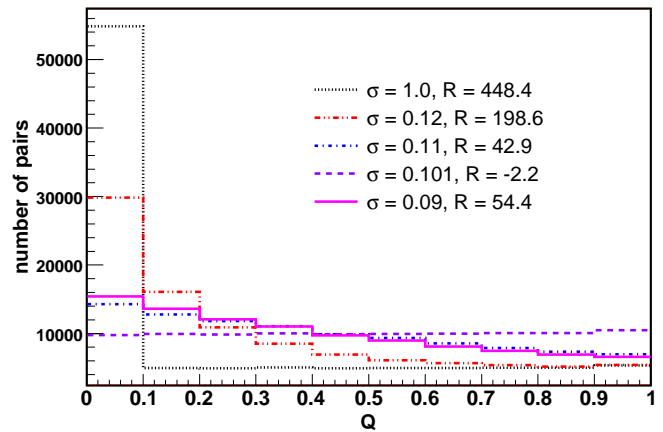}
\end{center}
\caption{
The $Q$-histograms from event samples consisting of two classes of events where 
rapidities were generated from Gaussian profiles with the same mean and the multiplicity 
was distributed Poissonian with the mean 512. The width in one class of events was 0.1. The 
width in the second class has been varied; different histograms correspond to different widths.
The values of $R$ are written in the legend.}
\label{f:varwidth}
\end{figure}

In Figure~\ref{f:varwidth}
we explore the effect of a variable width. One half of events was simulated with Gaussian distribution 
with the width of 0.1 and the other one with the same mean but a different width. The widths are
1.0, 0.12, 0.11, 0.101, and 0.09. The multiplicity was Poissonian-distributed with the mean of 512. 
We observe that except for the cases where the widths differ by  ten per cent or less the difference 
is picked up by the procedure. 

\begin{figure}[t]
\begin{center}
\includegraphics[width=1.0\linewidth]{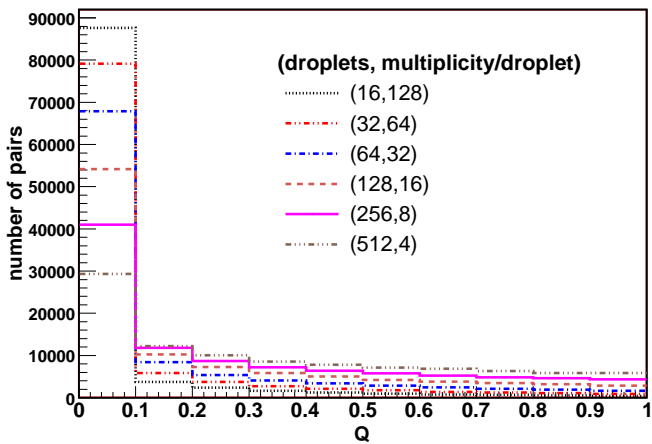}
\end{center}
\caption{
The effect of the number of Gaussian sources on the $Q$-histogram. In the legend, left is the 
number of Gaussian sources distributed uniformly between -1 and 1, right is the average number of 
pions from each source. The width of each Gaussian source was 0.707.}
\label{f:toydrop}
\end{figure}

Finally, we test a case which is closest to the fireball fragmentation scenario that we want to explore in detail. 
In Figure~\ref{f:toydrop}
we show the results from a simulation, where each event consists from superposition of many Gaussian distributions.
The width of all these distributions is 0.707 and is motivated by the typical rapidity spread of the pion rapidity 
at a realistic
freeze-out temperature. The means of the Gaussians are generated from a uniform distribution between --1 and 1. 
We test cases with 16, 32, 64, 128, 256, and 512 Gaussians per event, which emit on average 128, 64, 32, 16, 8, 4 
particles per Gaussian, respectively, so that the total multiplicity is always  2048. We observe, that even 
in the least favorable simulation with a large number of small droplets, the difference between events is 
clearly visible.


\section{Monte-Carlo droplet generator}
\label{quag}

Realistic events samples on which the KS test are applied were generated with the help of the 
Monte Carlo event genarator DRAGON \cite{Tomasik:2008fq}. Here we provide very brief overview 
of its capabilities. 

DRAGON assumes that the fireball decays into droplets which are distributed according to the blast-wave 
model. Thus their distribution in position and velocity is given by
\begin{equation}
S_D(x,v) \propto H(\eta)\,  \Theta(R-r)\,  \delta(\tau - \tau_0)\, \delta^{(4)}(v - u(x))\,  , 
\end{equation}
where we use polar coordinates $r$ and $\phi$, the space-time rapidity and longitudinal proper time
\begin{eqnarray}
\eta & = & \frac{1}{2} \ln \frac{t+z}{t-z} \\
\tau & = & \sqrt{t^2 - z^2}\, ,
\end{eqnarray}
as coordinates in the space-time. The fireball has a transverse radius $R$ and $\tau_0$ 
is the Bjorken proper time of the decay. The four-velocity of the droplet $v$ is given by the 
local flow velocity at the position where the droplet is created,
\begin{multline}
u_\mu(x) = (\cosh\eta\, \cosh\eta_t,\, \cos\phi\, \sinh\eta_t,\\
\sin\phi\, \sinh\eta_t,\, \sinh\eta\, \cosh\eta_t)\, ,
\end{multline}
with 
\begin{equation}
\eta_t  =  \frac{\sqrt{2}\rho_0 r}{R}\, , 
\end{equation}
where $\rho_0$ is a model parameter. (The model is designed so that it can simulate 
azimuthally non-symmetric fireballs, but we do not explore such a possibility here.)
The function $H(\eta)$ specifies the space-time rapidity distribution. It can be 
uniform or Gaussian. For the present investigation we use the uniform distribution
in rapidity.

The volumes of the droplets are random according to a gamma 
distribution
\begin{equation}
\label{gammadist}
{\cal P}_2(V) = \frac{1}{b\Gamma(2)}\,  \frac{V}{b} \, \exp\left (-V/b\right )\, ,
\end{equation}
with a model parameter $b$. The droplets decay into hadrons exponentially 
in time, so the times of emission of the droplets are distributed in the 
rest frame of the emitting droplet according to $\exp(-\tau/R_D)$, where 
$R_D$ is the radius of the droplet. A droplet emits hadrons according to thermal 
distribution with a temperature $T_k$, until it uses
up all of its mass. 
The mass of the droplet is determined according to its volume and the energy density
which is set to $0.7\, \mbox{GeV fm}^{-3}$. 

Hadrons may be emitted from the droplets or produced in the remaining space 
between them. The relative abundance of those emitted from droplets is specified as a model 
parameter. Hadrons emitted from the bulk are generated according to the blast-wave
emission function \cite{Schnedermann:1993ws,Csorgo:1995bi,Retiere:2003kf}
\begin{multline}
\label{emisf}
S(x,p)\, d^4x = \frac{2s+1}{(2\pi)^3}\,m_t \, \cosh(y-\eta)\,
\exp\left( - \frac{p^\mu u_\mu}{T_k}\right )\, \\
\times\Theta(R-r)\, H(\eta) 
     \,\delta(\tau - \tau_0)d\tau\, \tau\, d\eta\, r\, dr\, d\phi\, .
\end{multline}
Here the factor $(2s+1)$ denotes spin degeneracy. 

Resonances are included in the simulation. 
They decay according to the standard two-body or three-body
kinematics. Probabilities of production of individual species are given by the 
statistical model
with a chemical freeze-out temperature $T_{ch}$ and chemical potentials for baryon number 
and strangeness.


\section{Fluctuating rapidity distributions}
\label{results}

The Monte Carlo event generator DRAGON is employed \cite{Tomasik:2008fq} 
to simulate realistic data on which the KS test is performed. 
We use the test on data 
generated for RHIC  Au+Au collisions at $\sqrt{s}=200\,A$GeV and
FAIR Au+Au at $\sqrt{s} = 7.6\,A$GeV.
For the data analysis we accept hadrons within the rapidity interval 
[--0.5,0.5]. 

For RHIC, we have generated events with uniform rapidity distribution 
in the interval [--3,3]. The total hadron multiplicity was set to $dN/dy = 1000$. The chemical composition is 
determined by the following choice of parameters: $T_{ch} = 155$~MeV, $\mu_B = 26$~MeV
\cite{Andronic:2005yp}. We neglect the strangeness chemical potential. The list of 
resonances includes mesons up to a mass of 1.5~GeV/$c^2$ and baryons up to 2~GeV/$c^2$. 
The geometry of the decaying fireball is given by the radius $R=10$~fm and $\tau_0 = 9$~fm/$c$.
The dynamical state of the fireball is set by the kinetic freeze-out temperature $T_k = 150$~MeV
and the transverse expansion gradient $\eta_f = 0.6$. We  set the volume parameter of the 
droplets $b$ to the value of 10~fm$^3$. As a first benchmark, complementary samples of 10,000 events are 
generated: one with all particles being emitted from droplets, the other with all particles being 
emitted from the bulk fireball. 

As a second benchmark test we  generate 10,000 events at the FAIR energy of $\sqrt{s} = 7.6\,A$~GeV
where no particles are emitted from droplets. 
In this case the chemical freeze-out parameters are set to the corresponding values
$T_{ch} = 140$~MeV, $\mu_B=  375$~MeV, and $\mu_S = -53$~MeV. The kinetic freeze-out 
temperature is $T_k = 140$~MeV and the transverse expansion of the fireball is characterised 
by $\eta_f = 0.4$. Here, the rapidity distribution is Gaussian with a width of 0.7 and the total hadron 
multiplicity is 1,500. The transverse radius of the fireball and its Bjorken lifetime are
9~fm and 8~fm/$c$, respectively.

\begin{figure*}[t]
\begin{center}
\begin{minipage}{0.4\linewidth}
\includegraphics[width=1\linewidth]{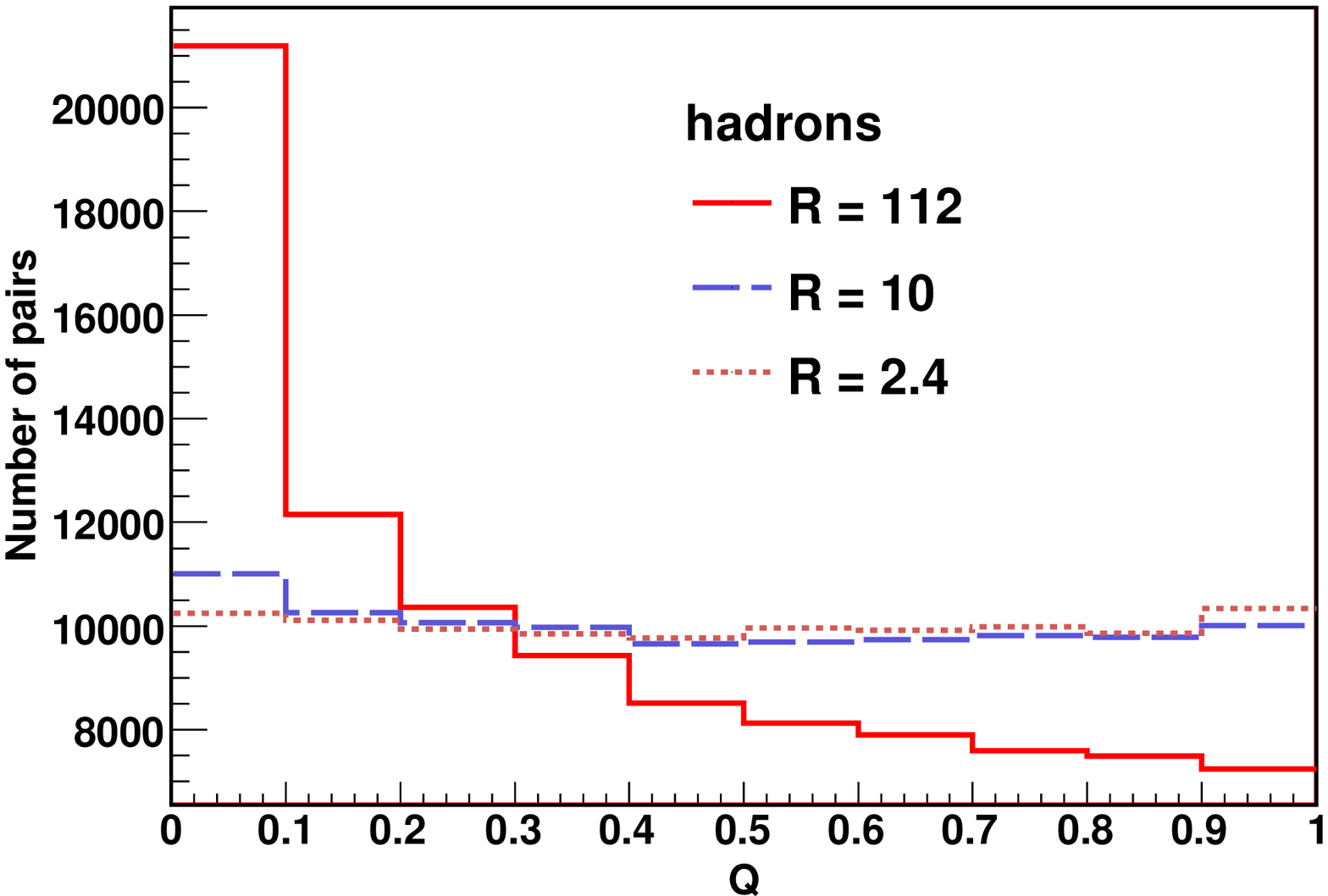}
\end{minipage}
\begin{minipage}{0.4\linewidth}
\includegraphics[width=1\linewidth]{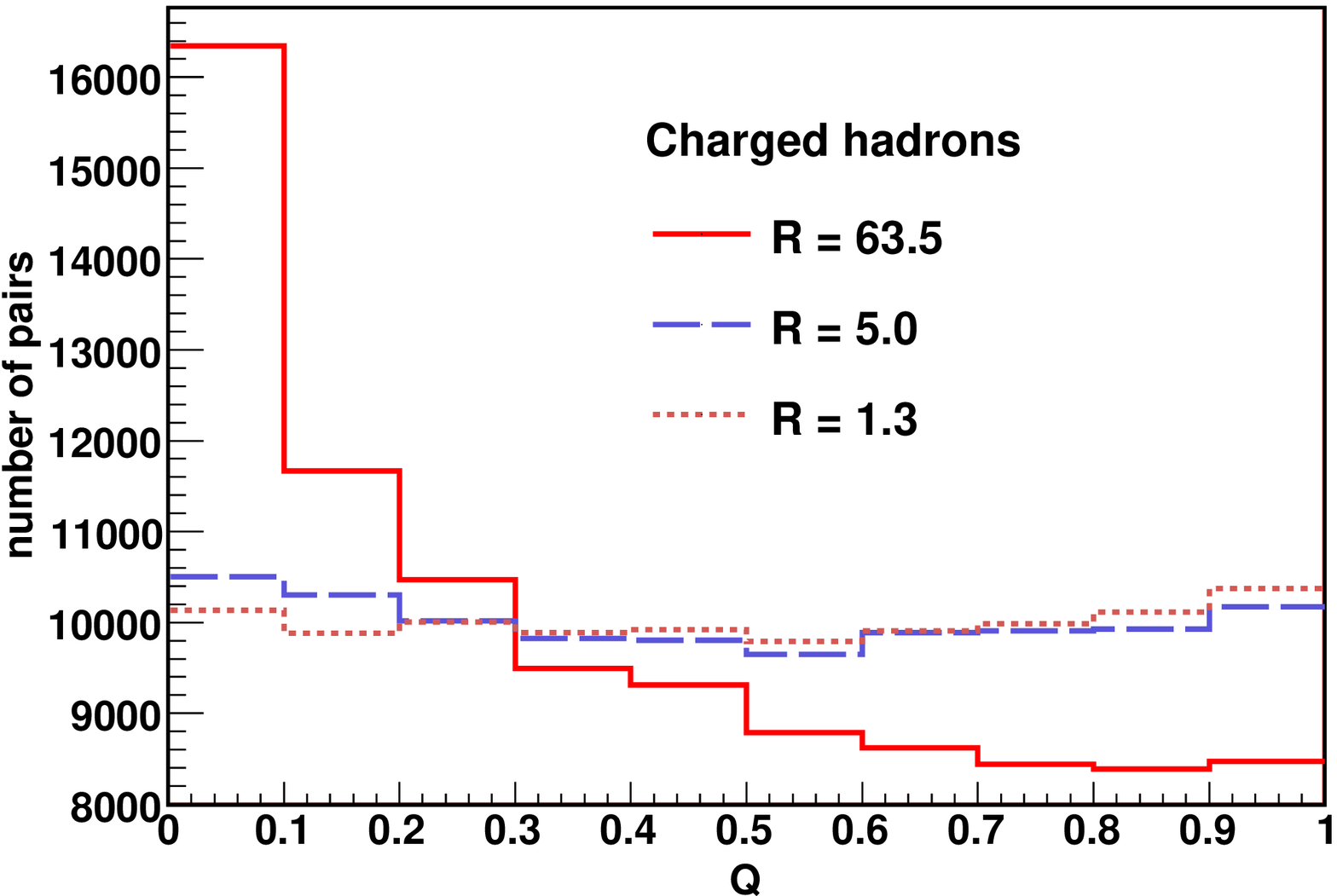}
\end{minipage}
\begin{minipage}{0.4\linewidth}
\includegraphics[width=1\linewidth]{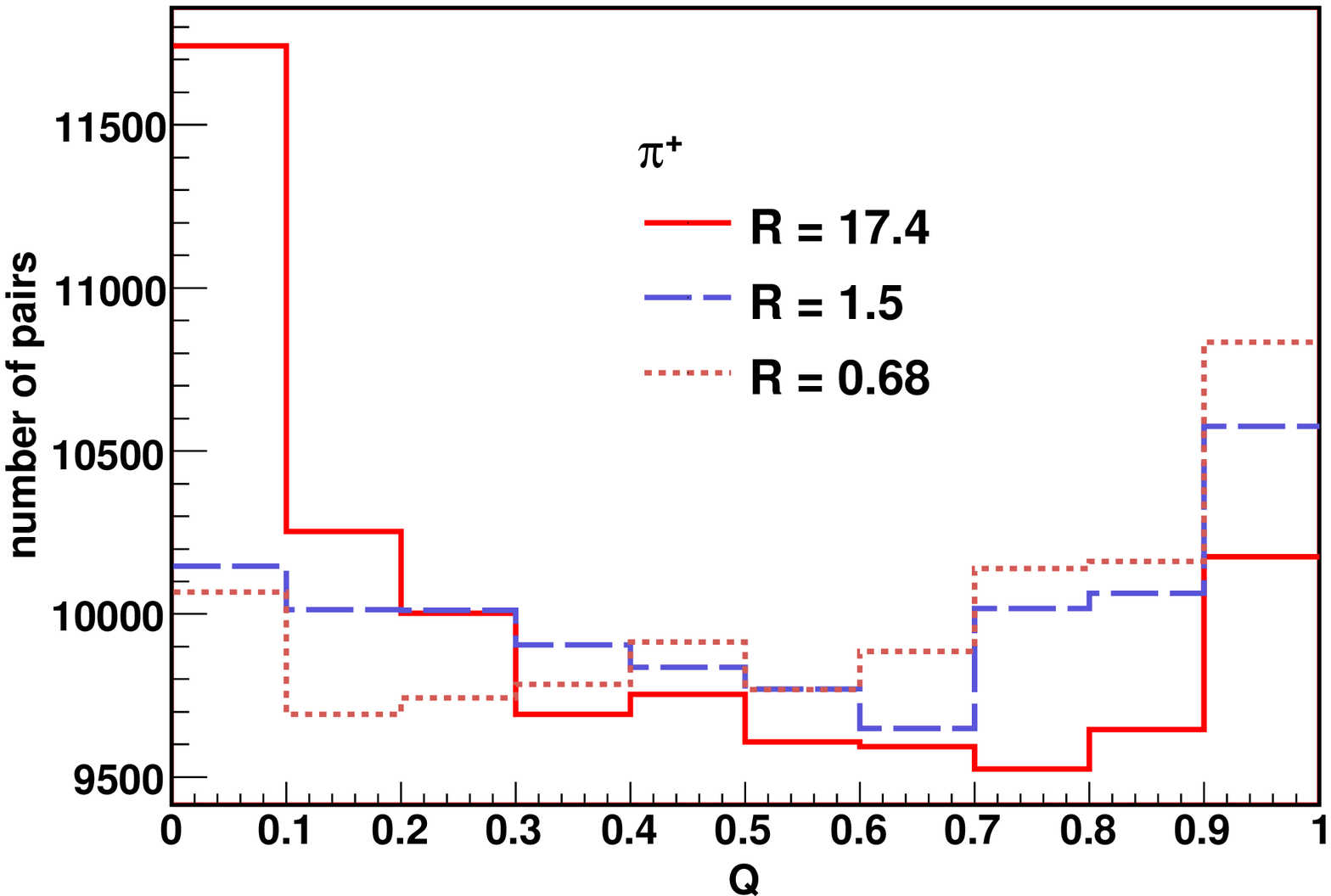}
\end{minipage}
\begin{minipage}{0.4\linewidth}
\includegraphics[width=1\linewidth]{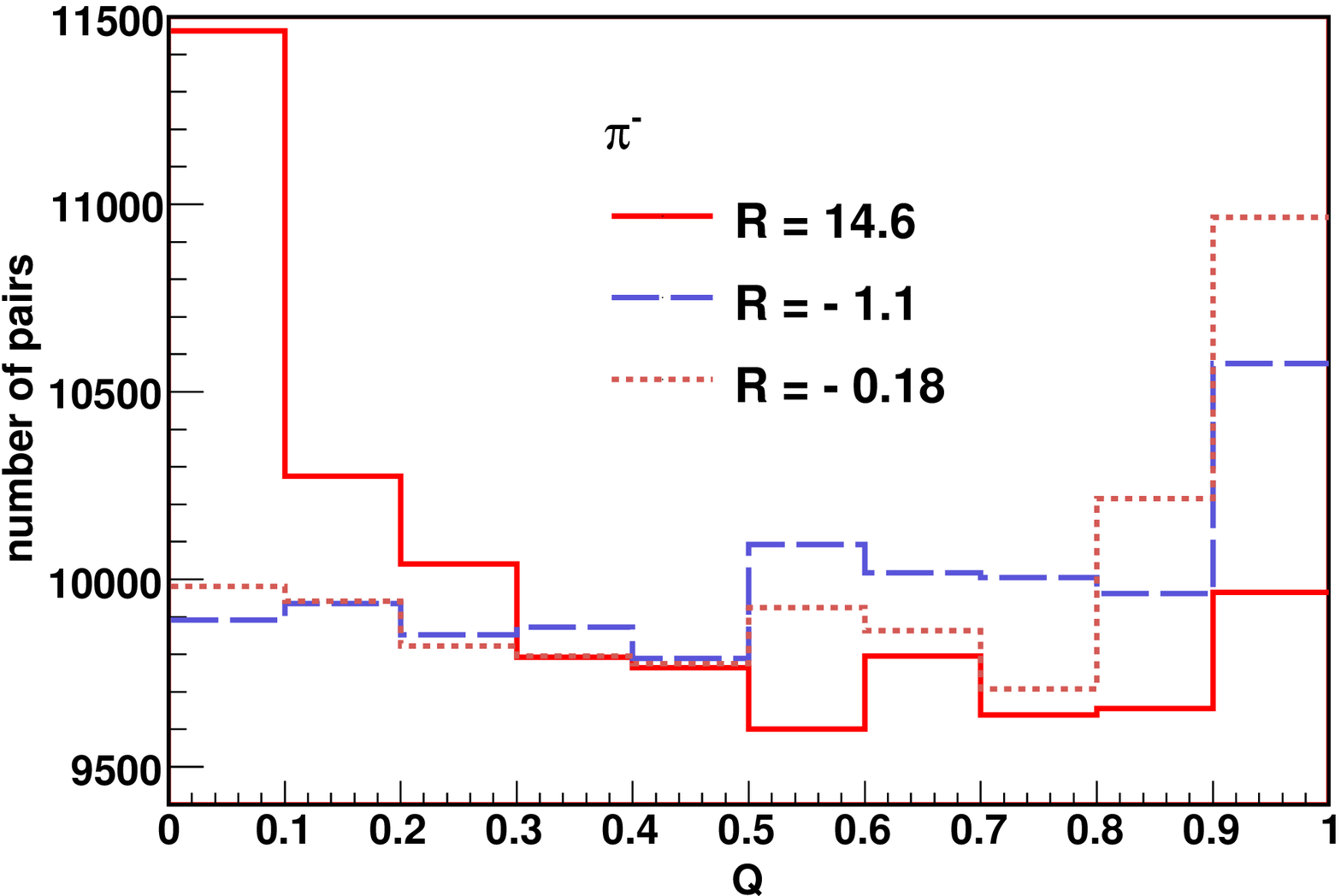}
\end{minipage}
\begin{minipage}{0.4\linewidth}
\includegraphics[width=1\linewidth]{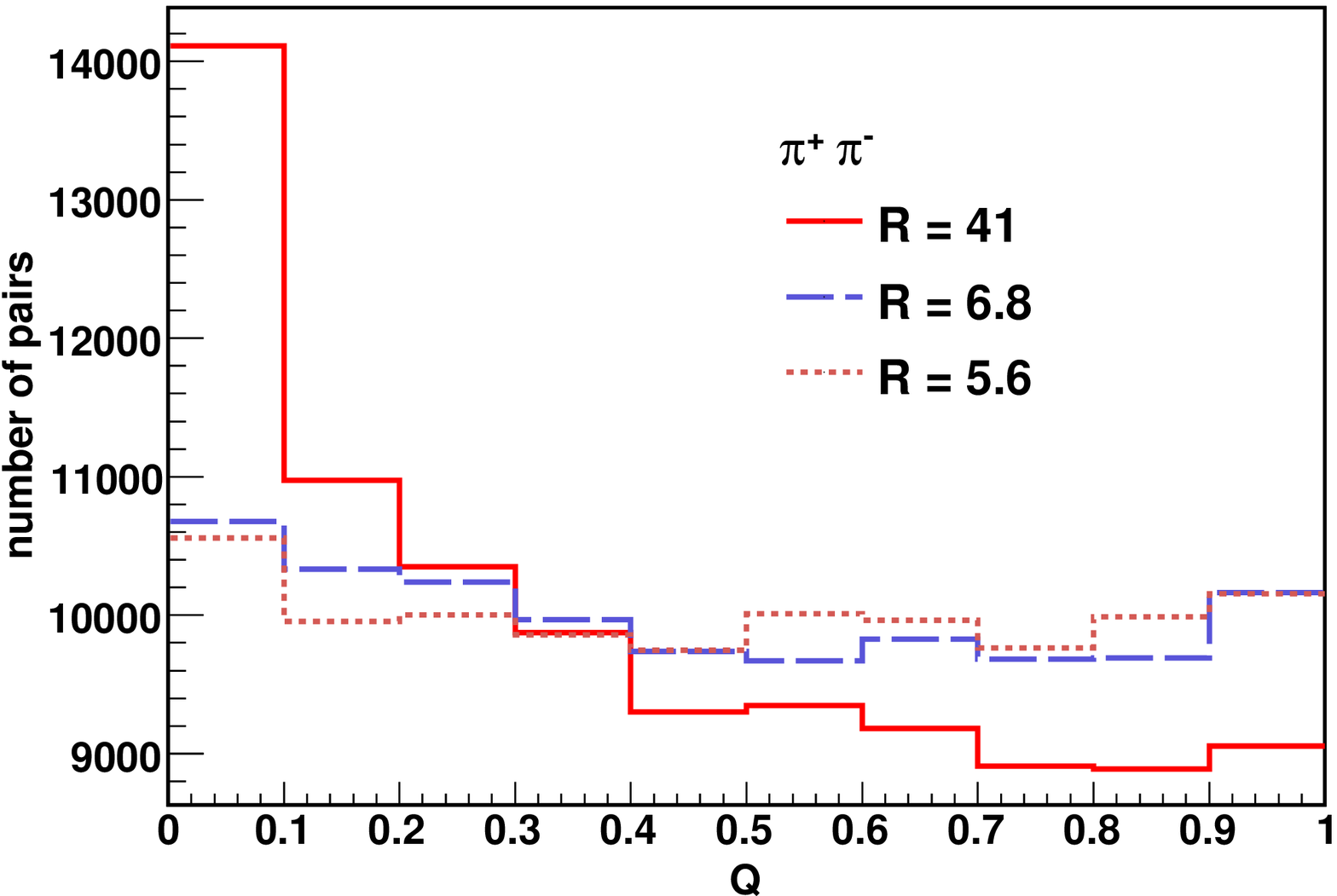}
\end{minipage}
\begin{minipage}{0.4\linewidth}
\includegraphics[width=1\linewidth]{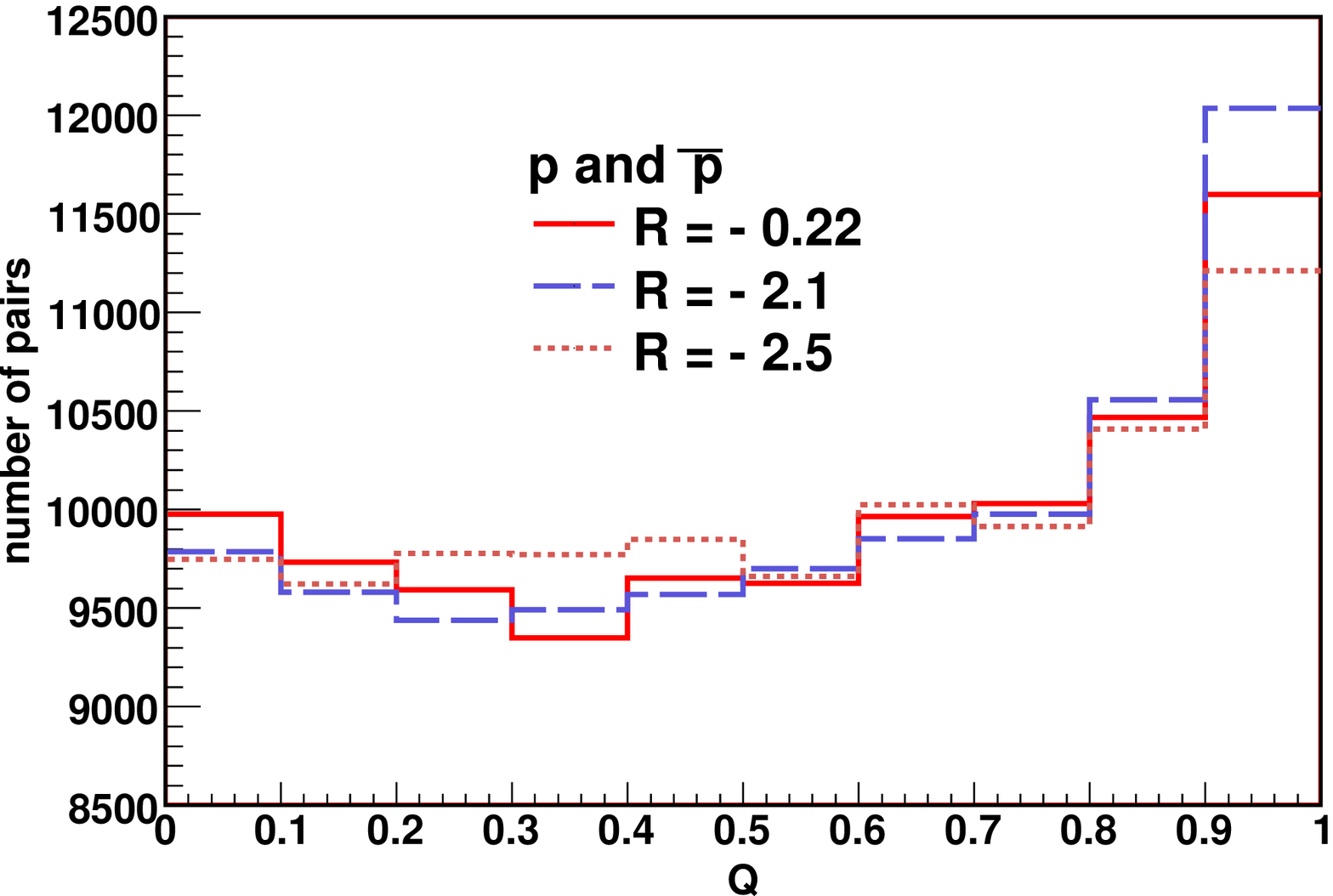}
\end{minipage}
\end{center}
\caption{
(Color online)
The $Q$-histograms resulting from simulations  of realistic hadronic final states with the help of DRAGON. 
Solid (red)  histograms correspond to simulation of RHIC Au+Au collisions with droplets. Dashed (blue)
histograms are from simulations for RHIC without droplets. Dotted (brown) histograms show the
results of simulations for nuclear collision at FAIR without fragmentation. Different panels show results 
obtained for all hadrons, charged hadrons, $\pi^+$, $\pi^-$, charged pions, protons and antiprotons.
The values of $R$ are indicated in the panels. 
}
\label{f:simpquag}
\end{figure*}

In figure~\ref{f:simpquag} we show the difference between the $Q$-histograms from 
events with and without droplets. 
For the RHIC energy, one observes a characteristic enhancement towards small $Q$ values
in the case of particle emission from droplets for all investigated particle species except
(anti-)protons. (For the setting without droplets (i.e.\ only bulk emission) this low $Q$ 
enhancement is strongly suppressed. Quantitatively, this is reflected in a factor of 10 difference
of the extracted $R$ values. The RHIC results without droplet formation are also in line with 
the results obtained at FAIR energies, showing that the KS test does not produce falsly positive 
results when going to smaller samples with a different rapidity distribution.

Resonance decays also have a clustering effect on the decay products. Therefore, a signal of clustering 
is also seen in the set of events without droplets. In case of all hadrons, these are mainly $\rho$'s and 
$\Delta$'s. If we limit our analysis to pions only, then there is correlation due to the $\rho$.
To test this hypothesis, one can perform the 
KS test with protons only, since there is no resonance that would decay into two baryons. The drawback
of using protons only is limited statistics in two ways. Firstly, their total multiplicity is lower,
e.g.\ there are only 10 to 50 protons in the acceptance per RHIC event.
Therefore, one observes fewer pairs at small $Q$ and a peak at $Q$ close to 1 
in case there are no droplets (see appendix).  
Secondly, if the droplets are small, protons are a less ideal probe because a droplet may not 
have enough energy to emit more than one proton and the correlation is gone then. 
An alternative solution is to use pions of the same charge. These are more abundant than protons
and no strong effect of resonances is seen here. Note, however, that we have not included the effect 
of pair wave function symmetrisation which leads to Bose-Einstein correlations. 


Note also, that resonances not only introduce correlations, they can also weaken the correlations 
due to droplets. Resonance decay products obtain some momentum due to higher mass
of the mother resonance. Thus, the velocities of decay products will be more smeared around the 
velocity of the droplet which emitted the resonance than the velocities of hadrons emitted from 
the droplet directly.

\begin{figure}[t]
\begin{center}
\includegraphics[width=1.0\linewidth]{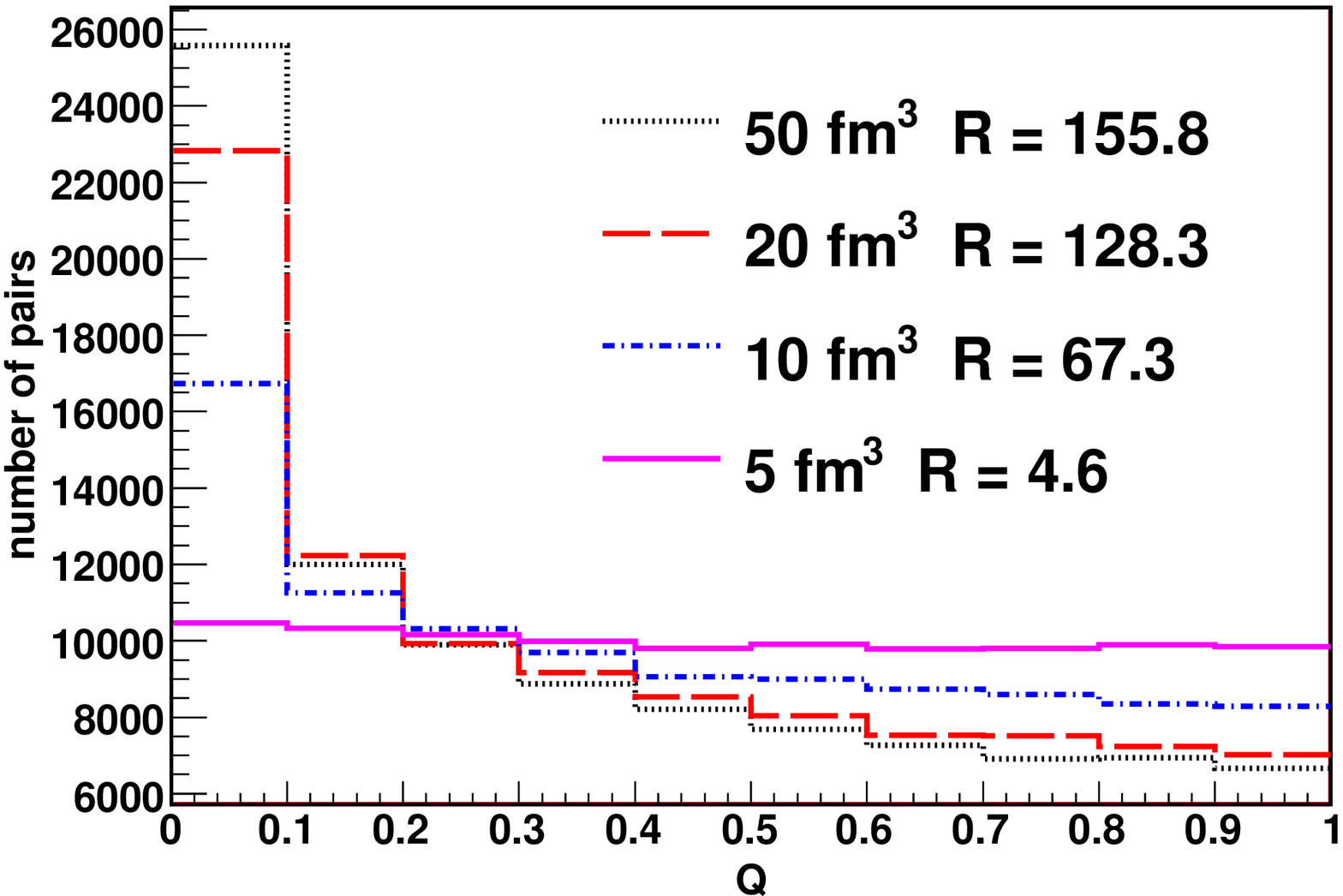}
\end{center}
\caption{
$Q$-histograms corresponding to various sizes of 
droplets: $b = 5,\, 10,\, 20,\, 50$~fm$^3$. The values of $R$ are shown in the 
legend.
The $Q$-histograms are obtained from samples of 10,000 events where all hadrons have 
been emitted from droplets. Only charged hadrons have been taken in constructing these histograms. 
}
\label{f:drsize}
\end{figure}

The influence of the size of the droplets is studied in Figure~\ref{f:drsize}.
The parameters of the simulation are kept the same as in the previous case, but the 
volume parameter $b$  varied to values 5, 10, 20, 50~fm$^3$. All 
particles are emitted from droplets. We do the KS test with 
charged hadrons. 
As expected from previous analysis (with $b=10\, \mbox{fm}^3$),
a dominant low $Q$ peak emerges for all droplet volumes down 
to 5~fm$^3$. From this we conclude that even small size droplets
can be detected with the analysis method presented here. 

As a final physics benchmark of the KS test we explore the effect 
of changing droplet fraction of the total multiplicity. A systematic 
study of how
the percentage of hadrons emitted from droplets affects the result is presented  
\begin{figure}[t]
\begin{center}
\includegraphics[width=1.0\linewidth]{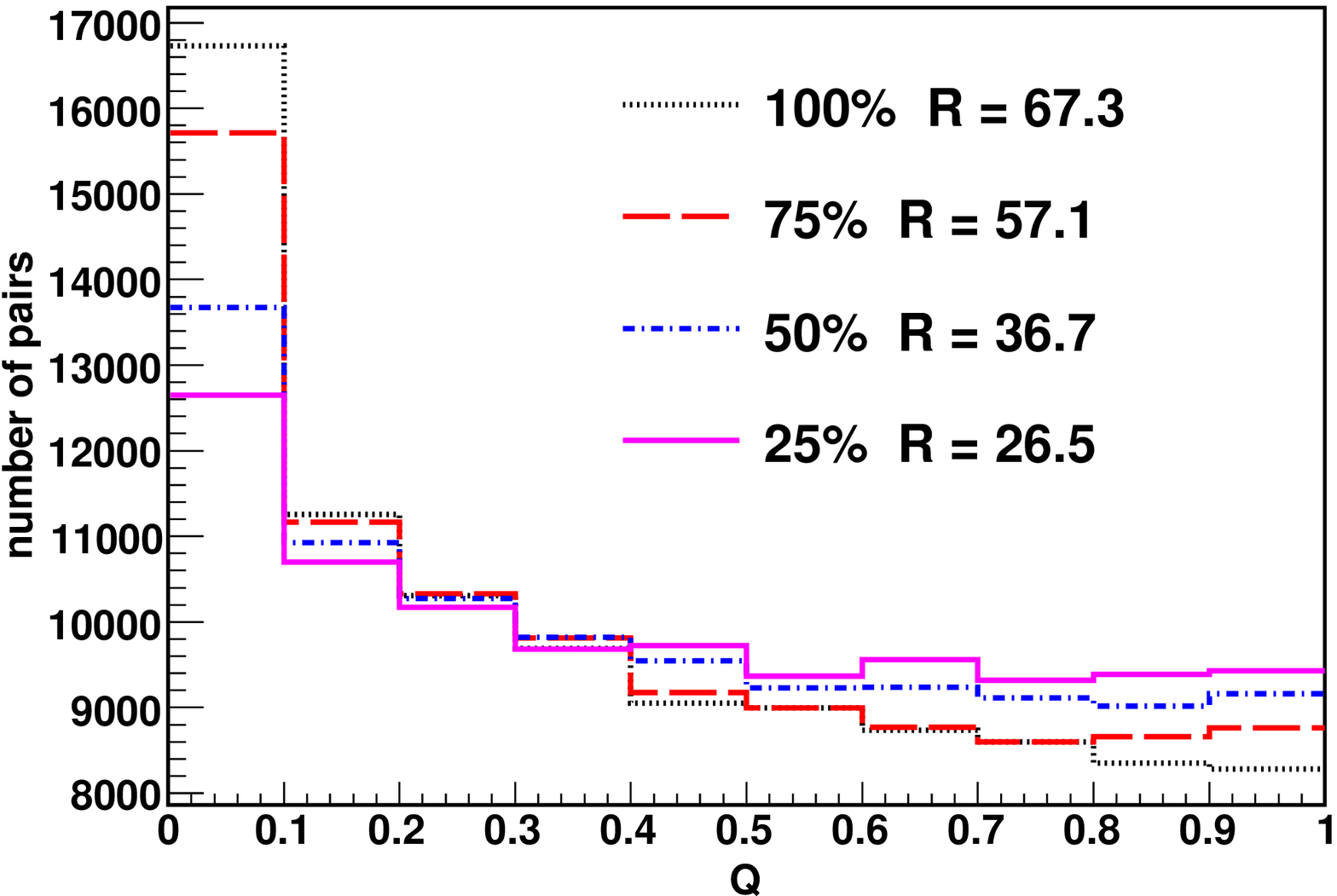}
\end{center}
\caption{
$Q$-histograms obtained from samples of 10,000 events where droplets with 
a volume parameter of
$b=10\, \mbox{fm}^3$ are present. The percentage of hadrons coming from droplets is 
varied: 25\%, 50\%, 75\%, and 100\%. The histograms are constructed with charged 
hadrons only. The values of $R$ are listed in the legend.}
\label{f:per}
\end{figure}
in Figure~\ref{f:per}. Here, the size of the droplets is fixed to 
$b = 10\, \mbox{fm}^3$ and  
only the percentage of hadrons from droplets is varied. Even if only one
quarter of all hadrons comes from the droplets and the rest from the 
gas in between them, the signal is well visible and the KS test can discriminate 
between the formation and the non-formation of droplets.


\section{Resolution}

\begin{figure}[t]
\begin{center}
\includegraphics[width=1.0\linewidth]{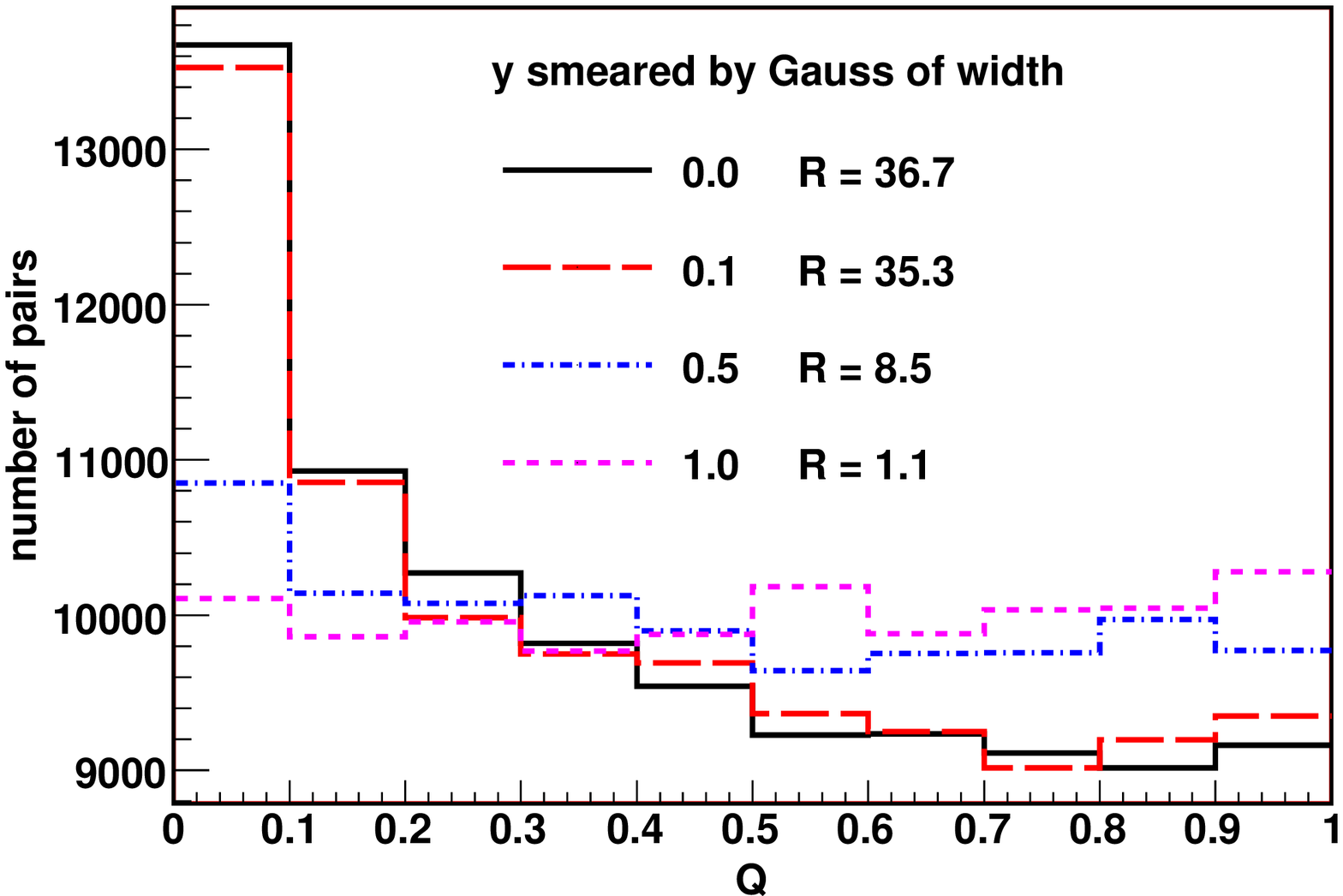}
\end{center}
\caption{
The influence of finite rapidity resolution. Charged hadrons were generated with the Monte 
Carlo event genarator DRAGON with $b=10\, \mbox{fm}^3$ and 50\% of hadrons generated
from droplets. Before the analysis, rapidities were smeared with Gaussian distribution
with the width 0.1 (long-dashed line), 0.5 (dash-dotted), 1.0 (short-dashed). Solid line
shows the result with non-smeared data. 
}
\label{f:reso}
\end{figure}

Finally, we address the experimentally
crucial question whether the droplet signal in the $Q$-histogram stays recognisable
if the rapidities are measured with finite resolution. We investigate this issue 
in Figure~\ref{f:reso}.
Events simulated with DRAGON for a volume parameter $b = 10\, \mbox{fm}^3$ and 
with 50 per cent of hadrons emitted 
from droplets are taken and the generated rapidities are smeared 
by a Gaussian with widths of 0.1, 0.5, and 1, to mimick the  finite resolution 
of experimental measurements. The events with smeared rapidities are then processed with the 
KS test. We observe a gradual weakening of the signal strength at low $Q$. For 
smearing by 0.5 units of rapidity the peak height becomes comparable with the peak 
resulting from resonance decays only (cf.\ Figure~\ref{f:simpquag}). 
For even poorer resolution, the peak can not be regarded as an unambiguous signal for 
droplet formation. The resolution, however, is usually on the level of $\Delta y \approx 0.1$.


\section{Conclusions}
\label{conc}

The Kolmogorov-Smirnov test is a powerful tool in searching for non-statistical 
differences between  events. The test itself is more general than investigated 
here, and applications will be presented in following papers. 
The logic of its use is the following: select a class of events
which are ``as identical as possible'', in particular in a  very narrow centrality class. 
Conventional scenarios predict that each event would evolve according to the same 
scenario and the final distributions of hadrons would be identical  in all events. 
The KS test is able to detect deviations from this scenario. If an effect is  observed, 
it remains to be studied
what phenomenon leads to positive results.

As a currently widely discussed topic we focussed the present investigation 
on the possible decay of the fireball into smaller droplets. The present study 
showed that the KS test is perfectly suited for this task and allows to extract a prominent 
signal. The signal is robust  even if only a small amount  
of the hadrons come from the droplets and survives realistic final rapidity resolution.
Thus, the test can be also used in a negative way: if its application on data yields 
only limited or no signature of non-statistical event-by-event fluctuations of the 
rapidity distributions, this puts limits on the scenarios assuming fireball fragmentation. 

The investigation of the signal of other effects (including a comparison to full Monte Carlo
transport simulation) in the KS test deserves separate studies 
and shall be performed in subsequent papers.


\acknowledgments

The work of BT, IM, SK, and MG has been supported by VEGA 1/4012/07.  The work of GT, SV, and MB 
was (financially) supported by the Helmholtz International
Center for FAIR within the framework of the LOEWE program
(Landesoffensive zur Entwicklung Wissenschaftlich-\"okonomischer
Exzellenz) launched by the State of Hesse.
BT acknowledges support from 
MSM~6840770039 and  LC~07048 (Czech Republic).
We thank Drs.~J.R.~Brown and M.E.~Harvey for valuable  
discussions.


\appendix

\section{Evaluation of the Kolmogorov-Smirnov distribution}

Throughout this paper we use the \emph{two-sample two-sided} 
(Kolmogorov-)Smirnov test\footnote{The Kolmogorov \emph{one-sample}
test refers to a comparison of one empirical cummulative distribution 
function based on data with a smooth thoretical distribution function.
This is distinguished from the two-sample (Smirnov) test where two data samples 
are compared with each other. One-sided and two-sided tests refer 
simply to the difference of two cumulative distribution functions 
or to its absolute value, respectively.}.
The cummulative distribution function of the difference $D$ for the
one-sample test in case $n\to \infty$ was derived by Kolmogorov \cite{kolm}:
\begin{equation}
\label{k0}
Q(D) = K_0(D) = -2 \sum_{k=1}^{\infty} (-1)^k \exp\left ( -2 k^2 D^2 \right )\, .
\end{equation}
Later, Smirnov \cite{smir} proved that the same result applies for the two-sample test for 
$n_1,n_2 \to \infty$ under replacement $D\to d = D\sqrt{n_1 n_2/(n_1 + n_2)}$. 
We have checked that such an asymptotic 
case is not a good approximation even for $n$'s around 200. It is therefore desirable to obtain 
formulas valid in non-asymptotic case. 

A simple solution is to replace the quantity $d = \sqrt{n} D$ in eq.~\eqref{k0} with the following one, 
originally due to Stephens 
\cite{Stephens} (found also in Numerical Recipes \cite{nr}):
\begin{equation}
\label{NR}
d = D \Big(\sqrt{n} + 0.12 + \frac{0.11}{\sqrt{n}}\Big)
\end{equation}
A different formula with
a few terms of an expansion in $1/\sqrt{n}$ has been
derived by Li-Chien \cite{chang,peltz}. In such an expansion 
\begin{equation}
\label{k123}
Q(d) = K_0(d) + K_1(d) + K_2(d) + K_3(d) + \dots \, .
\end{equation}
The leading order term has been displayed in eq.~\eqref{k0}. The following three terms are
\begin{widetext}
\begin{eqnarray}
\label{k1}
K_1(d) & = & \frac{4 d}{3 \sqrt{n}} \sum_{k=1}^{\infty} (-1)^k k^2 \exp\left ( -2 k^2 d^2 \right )\\
\label{k2}
K_2(d) & = & \frac{1}{9n} \sum_{k=1}^{\infty} (-1)^k 
\Biggl (  k^2 - \frac{1}{2} \left (1 - (-1)^k\right )
				- 4k^2 d^2 \left ( k^2 - \frac{1}{2}\left ( 1 - (-1)^k \right ) + 3 \right )    
				+ 8 k^4 d^4  \Biggr )
\exp\left ( -2 k^2 d^2  \right ) \\
\nonumber
K_3(d) & = & - \frac{2 d}{27 n^{3/2}} \sum_{k = 1}^{\infty} (-1)^k k^2 
\Biggl (  \left ( k^2 + \frac{22}{5} - \frac{3}{2} \left ( 1 - (-1)^k  \right )     \right ) \\
& & \qquad \qquad \qquad \qquad \qquad {}
- k^2 d^2 \left (\frac{4}{3} k^2 + \frac{88}{15} - 2 \left ( 1 - (-1)^k  \right ) +12  \right )
+ 8 k^4 d^4 \Biggr )  \exp \left ( -2 k^2 d^2 \right )\, .
\label{k3}
\end{eqnarray}
\end{widetext}
These relations were derived for one-sample test. We checked, however, that they are much easier to 
handle and give better results when tested on samples of statistically identical events (see below) than 
approximations to two-sample distributions for non-asymptotic cases \cite{kim,brown}. Therefore, we
decided to use these relations although we note that a revision of the formulae for two-sample tests 
is desirable. 

\begin{figure}[t]
\begin{center}
\includegraphics[width=1.0\linewidth]{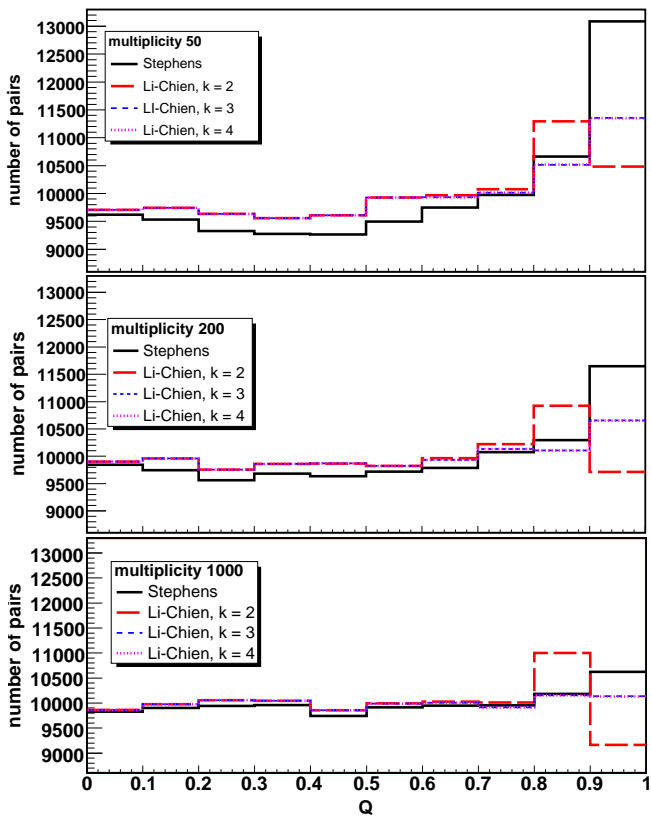}
\end{center}
\caption{Stephens' approximation according to eq.~\eqref{NR} vs. Li-Chien approximation (eq.~\eqref{k123}) 
with varying number of terms after which it is truncated. The short-dashed blue line coincides with the dotted purple line.}
\label{fig:Poiss(50)}
\end{figure}

In practical calculations it turns out that it is sufficient to cut off the
expansions in eqs.~\eqref{k0}, (\ref{k1}--\ref{k3}) at $k = 4$. To illustrate this point, we compare in 
Fig.~\ref{fig:Poiss(50)} 
the histograms of $Q$'s calculated from  eq.~\eqref{NR} with histograms 
based on eq.~\eqref{k123} with the cut-off at $k=2,\, 3$ and 4 respectively.
We show the results for $10^5$ pairs of events chosen randomly out of $10^5$ simulated events. Each simulated event is 
represented by `rapidities' uniformly generated on (0,1) with Poissonian distributed
multiplicities with mean values of 50, 200 and 1000.
Thus we calculate $Q(d)$ using the expansion 
\eqref{k123} up to the term $K_3$ and evaluate the sums in eqs.~\eqref{k0}, (\ref{k1}--\ref{k3})
up to the fourth order in $k$. 

The Li-Chien approximation truncated after $k=4$ may lead to negative $Q$ for $d>1.94$. Such a pair 
would fall into the last $Q$-bin. To fix the problem  for this value of $d$ an approximation 
due to Marsaglia \cite{marsaglia} is employed
\begin{equation}
Q(d)  = 2 \exp \left [ \left ( -2.000071 - \frac{0.331}{\sqrt{n}} - \frac{1.409}{n} \right ) nd^2 \right ]\, .
\end{equation}


\end{document}